# Binary Mixtures of Novel Sulfoxides and Water: Intermolecular Structure, Dynamic Properties, Thermodynamics, and Cluster Analysis

Vitaly V. Chaban

PES, Vasilievsky Island, St. Petersburg, Russian Federation.

**Abstract**. Senior dialkyl sulfoxides constitute interest in the context of biomedical sciences due to their abilities to penetrate phospholipid bilayers, dissolve drugs, and serve as cryoprotectants. Intermolecular interactions with water, a paramount component of the living cell, determine performance the sulfoxide-based artificial systems in their prospective applications. Herein, we simulated a wide composition range of the sulfoxide/water mixtures, up to 85 w/w% sulfoxide using classical molecular dynamics to determine structure, dynamics, and thermodynamics as a function of the mixture composition. As found, both diethyl sulfoxide (DESO) and ethyl methyl sulfoxide (EMSO) are strongly miscible with water. DESO- and EMSO-based aqueous mixtures exhibit similar structure and thermodynamic properties, however, quite different dynamic properties over an entire range of compositions. Strong deviations from an ideal mixture between 30-50 mol% of sulfoxide content leads to relatively high shear viscosities of the mixtures. Free energy of mixing with water is only slightly more favorable for EMSO than for DESO. The results, for the first time, quantify high miscibilities of both sulfoxides with water and motivate comprehensive in vivo investigation of the proposed mixtures.

TOC

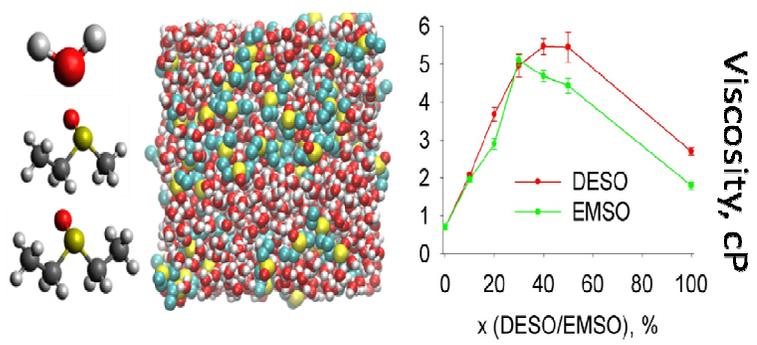

**Introduction**

Sulfoxides represent a small and rather peculiar group of organic compounds, in which sulfur exhibits valence of four. Thanks to a double sulfur-oxygen covalent bond in their structures, sulfoxides belong to rather polar compounds. Dialkyl sulfoxides are liquids at room conditions maintaining quite low shear viscosity values. The omnipresent antiparallel orientation of the sulfur-oxygen bonds is a major structural motif of the condensed dialkyl sulfoxides.[1-9] By possessing both polar oxygen and nonpolar side chains, sulfoxides exhibit complex intermolecular interactions with hydrophilic and lipophilic compounds.[3,10-14]

The most widely studied sulfoxide, dimethyl sulfoxide (DMSO), possesses outstanding solvation properties for medical and biological applications, being also often used in organic synthesis as a valuable, and in many senses unique, mild oxidant.[7,9,11,12,15] Furthermore, DMSO readily penetrates skin (phospholipid bilayers), dissolves drugs to obtain pharmaceutically relevant concentrations of the active substances, and efficiently transports them from outside. Well-known medicinal applications of DMSO are to decrease pain, speed up healing of wounds, treat muscle and skeletal injuries, recover burns, etc. Toxicity and long-term adverse effects of DMSO intoxication for human beings are extremely modest. More senior sulfoxides are much less extensively studied,[13,16-19] however, certain data suggests that they can be better cryoprotectants for biological research procedures. Fundamental interest to designing possible novel surface-active molecular agents (in the case of rather long alkyl chains) applies as well.

Markarian and coworkers have pioneered diethyl sulfoxide research in the 21$^{st}$ century, outlined a perspective biological context of research. Their numerous experimental results[13,14,16-18,20,21] advocate favorable cryoprotective properties of diethyl sulfoxide (DESO), thanks to the higher values of membrane potential and specific growth rate of E. coli. It must be noted that DESO can be synthesized with a great degree of purity, in which it provides cryopreservation of cells and tissues. DESO inhibits formation of large ice crystals making freezing of biological entities more sustainable, with a death rate decreased.[22] Mixtures of DESO with water exhibit strongly negative deviations from an ideal behavior.[13,16,18] Thanks to hydrophobic chains, DESO penetrates living cells somewhat better, as compared to DMSO, without causing significant damage to the bilayer. DESO also exhibits a different effect on the micelle formation of sodium dodecyl sulfate increasing the critical micellization concentration and changing water structure. Sulfoxides tend to decrease surface tension of water in their mixtures and exhibit an elevated density at the sulfoxide/water interfaces.[23]

Recently, Gabrielyan measured Fourier-transform infrared spectra for a novel sulfoxide, diisopropyl sulfoxide (DiPSO), in its pure state and binary mixtures with water and tetrachloromethane.[24] Quantum chemical calculations were used to describe an isolated DiPSO molecule and its complex with a water molecule in the ground state. DiPSO is an interesting organic solvent with an unusual structure of a hydrophobic chain.

Russina and coworkers[19] reported a versatile and highly insightful X-ray diffraction and molecular dynamics simulation study of liquid dibutyl sulfoxide (DBSO) at 320 K. DBSO exhibits an enhanced dipole–dipole correlation, as compared to the shorter chain sulfoxides, and tends to self-organize. In the meantime, behavior and described structural patterns of hydrophobic chains in DBSO resembles their role in lipid bilayers. New observations, such as structural differentiation, enhanced dipole-dipole correlation, limited role of hydrogen bonding proved experimentally,[19] specific mostly to medium-length amphiphilic sulfur-based organic molecules have been described suggesting a net value of future research in this direction. At this point, both experimental and theoretical data on senior dialkyl sulfoxides is insufficient to discuss their large-scale practical applications in lieu of or in conjunction with DMSO.

The present work provides a comprehensive investigation of dialkyl sulfoxide/water systems, for the first time discusses physical chemical properties of asymmetric sulfoxide, ethyl methyl sulfoxide (EMSO), in dilute and concentrated water solutions, correlates structure, transport, and standard free energies of hydration. An opposite situation, i.e. dilute water solutions in DESO and EMSO, has also been considered. This computational theoretical work particularly focuses on microscopic understanding of dialkyl sulfoxide – water interactions at the atomistic resolution at supposedly experimentally relevant compositions of their mixtures. The performed cluster analysis attends affinity of DESO and EMSO to self-association, a problem that has never been scrutinized thus far.

**Methodology**

A comprehensive set of physical chemical properties (thermodynamics, structure, dynamics) of the DESO-water and EMSO-water binary mixtures was derived from equilibrium trajectories (15-20 ns) of classical molecular dynamics (MD) simulations, except solvation Gibbs free energies, see below. The first 6.5 ns of each MD simulation were disregarded as a relaxation (equilibration) stage. The MD simulations were conducted in the constant pressure constant temperature ensemble, often denoted as NPT. The average constant temperature, 300 K, was maintained by the velocity rescaling thermostat[25] (relaxation time constant 0.5 ps), providing a correct distribution for a given ensemble, whereas the average constant pressure was maintained by the Parrinello-Rahman method[26] (relaxation time constant 2.0 ps). The Newton's equations of motion were integrated following the leap-frog algorithm with a time-step of 0.001 ps. Energy and pressure components during the equilibrium (productive) stage were saved every 0.015 ps and atomic coordinates were saved every 0.5 ps for subsequent derivation of physical chemical properties. Extent of sampling and frequency of data saving were thoroughly tested by using multiple shorter time periods and a point-to-point comparison of the corresponding results with those from longer time periods. It was found that the selected simulation details provide perfect sampling and exclude any systematic errors and unnecessary large standard deviations due to insufficient investigation of the system phase space.

The force field developed for DESO and EMSO recently was employed for MD simulations.[27] In turn, the water molecules were simulated by means of the SPC/E model, in its compatible (GROMOS96) implementation.[28] Electrostatic interactions were simulated directly (by the Coulomb law) if the distance between any two interacting centers was smaller than 0.9 nm. If at any point of time any interatomic distance exceeded 0.9 nm, the Particle-Ewald-Mesh technique was employed.[29] The Lennard-Jones (12, 6) short-range intermolecular atom-atom attraction was smoothly modified between 0.85 nm and the cut-off distance to vanish completely at 0.90 nm. These methodological parameters were pre-determined by the choice of the well previously tested force field. The list of neighboring atoms was updated every 12 time-steps within the radius of 0.9 nm to warrant reliable total energy conservation. The compositions of the simulated MD systems are given in Table 1. In view of the MD system sizes, the simulations were conducted in parallel using 8 cores per system in conjunction with cubic (2×2×2) domain decomposition scheme.

Table 1. Compositions, sizes, and sampling times of the simulated MD systems.

| # | Sulfoxide molar fraction, % | Mass fraction, % | # sulfoxide molecules | # water molecules | # interaction centers | Duration, ns |
|---|---|---|---|---|---|---|
| 1 | 0 | 0 | 0 | 1000 | 3000 | 5.0 |
| 2 | 10 | 40 | 100 DESO | 900 | 3300 | 20 |
| 3 | 20 | 60 | 200 DESO | 800 | 3600 | 20 |
| 4 | 30 | 72 | 300 DESO | 700 | 3900 | 20 |
| 5 | 40 | 80 | 400 DESO | 600 | 4200 | 20 |
| 6 | 50 | 86 | 500 DESO | 500 | 4500 | 20 |
| 7 | 100 | 100 | 1000 DESO | 0 | 6000 | 20 |
| 8 | 10 | 36 | 100 EMSO | 900 | 3200 | 15 |
| 9 | 20 | 56 | 200 EMSO | 800 | 3400 | 15 |
| 10 | 30 | 69 | 300 EMSO | 700 | 3600 | 15 |
| 11 | 40 | 77 | 400 EMSO | 600 | 3800 | 15 |
| 12 | 50 | 84 | 500 EMSO | 500 | 4000 | 15 |
| 13 | 100 | 100 | 1000 EMSO | 0 | 5000 | 15 |
| 14 | →0 | →0 | 1 DESO | 600 | 1806 | 30 |
| 15 | →0 | →0 | 1 EMSO | 600 | 1805 | 30 |
| 16 | →0 | →0 | 400 DESO | 1 | 2403 | 30 |
| 17 | →0 | →0 | 400 EMSO | 1 | 2003 | 30 |

Energetic and structure properties were computed using their conventional definitions. Self-diffusion coefficients were obtained from the mean-squared displacements (Einstein formula), shear viscosity was obtained via Green–Kubo integral of the pressure tensor autocorrelation function. A large number of trajectories were used for every MD system to derive reliable transport properties and their standard deviations. Gibbs free energies of solvation of DESO/EMSO in water and of water in DESO/EMSO were evaluated by thermodynamic integration. The solvated molecule was slowly and gradually removed (decoupled) from the solvent during 30 ns by decreasing solute-solvent interactions, whereas other interactions remained intact.

MD simulations were carried out in GROMACS 4 program.[30-32] Systems were visualized and the corresponding artwork (Figure 1) was prepared in the Visual Molecular Dynamics software (version 1.9.1).[33] Packmol was used to get initial configurations.[34]

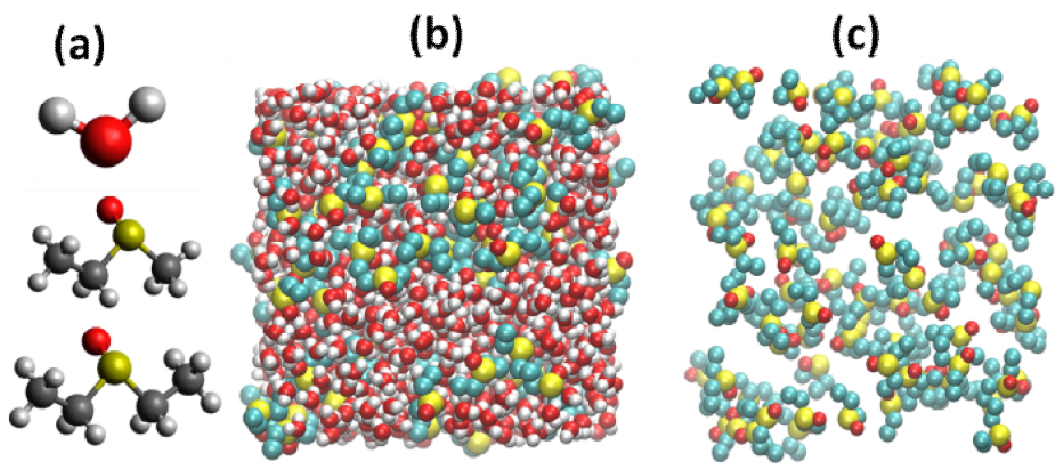

Figure 1. (a) Force-field optimized geometries of the isolated molecules constituting the simulated liquids: water, ethyl methyl sulfoxide, and diethyl sulfoxide; (b) equilibrated molecular configuration of the 10 mol% DESO mixture with water at 300 K and 1 bar; (c) 10 mol% DESO mixture with water molecules made invisible. Oxygen atoms are red, sulfur atoms are yellow, carbon atoms are grey, hydrogen atoms are white, united-atom interaction sites (-$CH_2$ and -$CH_3$) are cyan.

**Results and Discussion**

Figure 2 summarizes cohesive energies of the dialkyl sulfoxide/water mixtures as a function of DESO/EMSO content. Addition of sulfoxides to water decreases cohesive energy of the mixture more than threefold due to weaker sulfoxide – sulfoxide interaction, as compared to water-water interactions. In the meantime, water – sulfoxide non-covalent attraction remains quite strong, as exemplified in Figure 1. Indeed, DESO molecules are dispersed throughout the box, even with an account for their strong S=O self-association patterns. In one of the conducted simulations, DESO/EMSO molecules were initially placed separately from water molecules, i.e. mixing of the two liquids at 300 K was studied. No energy barriers corresponding to mixing were recorded. The total energy of the mixture was decreased to its target (equilibrium) value within a few nanoseconds. No specific sampling procedures were applied at this stage. Note that the difference of cohesive energies in DESO-rich and EMSO-rich aqueous mixtures is modest. It suggests that structure patterns in both compared systems are similar.

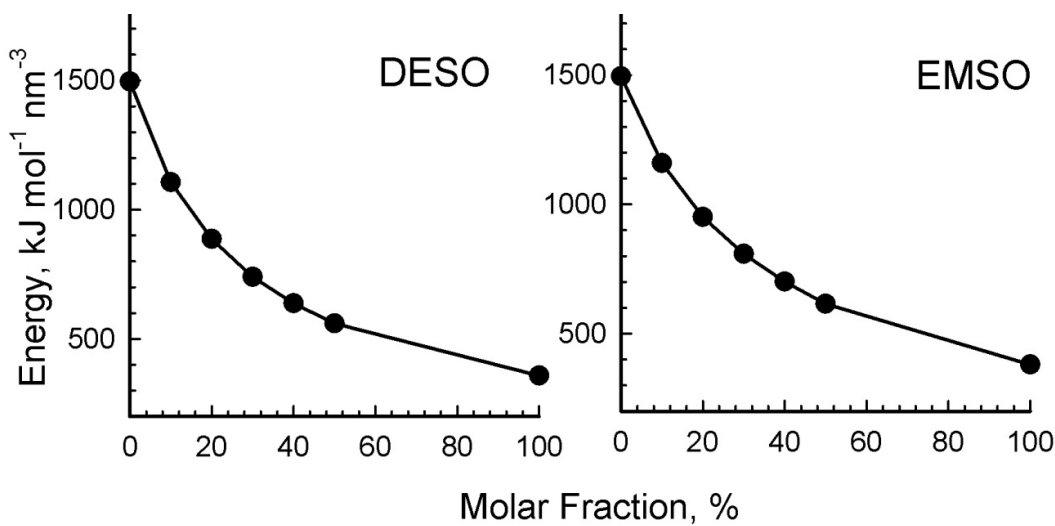

Figure 2. Cohesive energies of DESO (left) and EMSO (right) mixtures with water as a function of the DESO (EMSO) molar fraction at 300 K and 1 bar. The size of symbols is comparable to the error bars computed.

Sulfoxides are 5-6% denser liquids than water, even though their respective cohesive energies are clearly inferior to those of water. The different trends are due to relatively heavy sulfur atom, but not due to a closer distance between neighboring molecules. Addition of sulfoxides increases the density of all mixtures somewhat (Figure 3). Both DESO and EMSO exhibit very similar trends vs. molar fraction in the aqueous mixtures. The conducted simulations found no effect of the asymmetry of EMSO on thermodynamic properties of the mixtures, probably because the methyl chain is not involved in any strong intermolecular attraction.

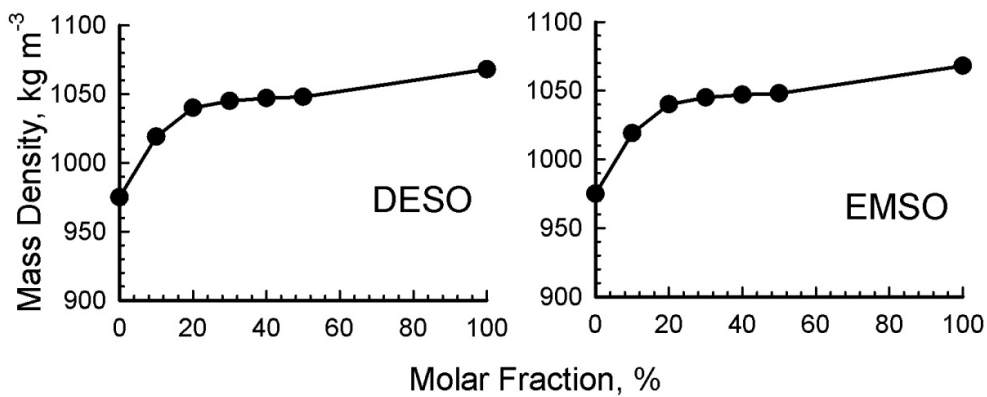

Figure 3. Mass density of DESO (left) and EMSO (right) mixtures with water as a function of the DESO (EMSO) molar fraction at 300 K and 1 bar. The size of symbols is comparable to the error bars computed.

Molar volumes (Figure 4) increase with an addition of sulfoxide molecules similarly to mass density (Figure 3). The deviation from ideality is moderately negative, i.e. sulfoxide/water mixtures are more compressed at the molecular level than pure liquids. An absence of disruptions in the plot indicates a good miscibility of these two polar liquids. Molar volume of pure DESO is only slightly larger than that of EMSO. One assumes that negative deviations decrease with an increase of the alkyl chain length and at some point become positive. For instance, the results of Russina and coworkers imply substantial degree of clustering in dibutyl sulfoxide.

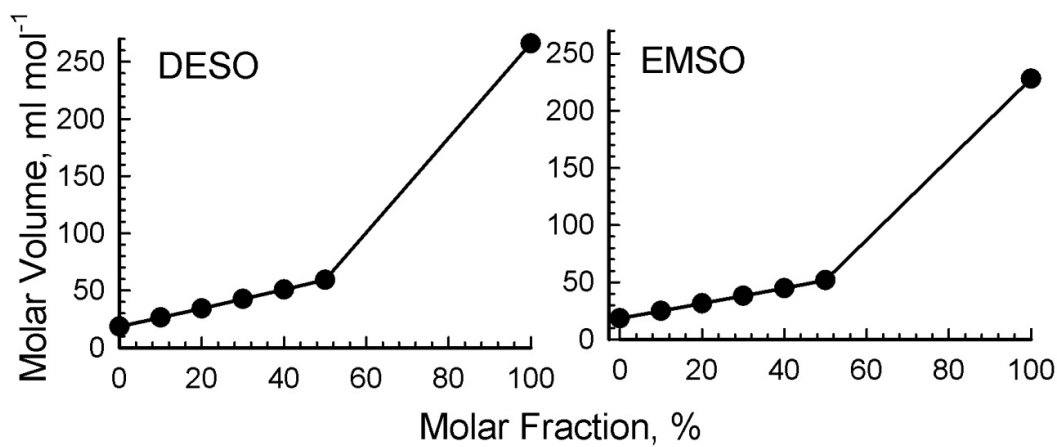

Figure 4. Molar volume of DESO (left) and EMSO (right) mixtures with water as a function of the DESO (EMSO) molar fraction at 300 K and 1 bar. The size of symbols is comparable to the error bars computed.

Radial distribution functions indicate a strong oxygen – sulfur structure correlation (2.7 units), 0.24-0.28 nm, also known as antiparallel dipole-dipole orientation (Figure 5). Another relatively strong structure correlation is oxygen-methyl pair (intermolecular) at 0.38-0.42 nm. Indeed, an interesting feature of dialkyl sulfoxides is their ability to engender a very weak hydrogen bond between electron-rich oxygen and the terminal methyl group. A strong hydrogen bond between oxygen of DESO/EMSO and water, 0.19-0.21 nm, is responsible for a good miscibility of both liquids over an entire composition range.

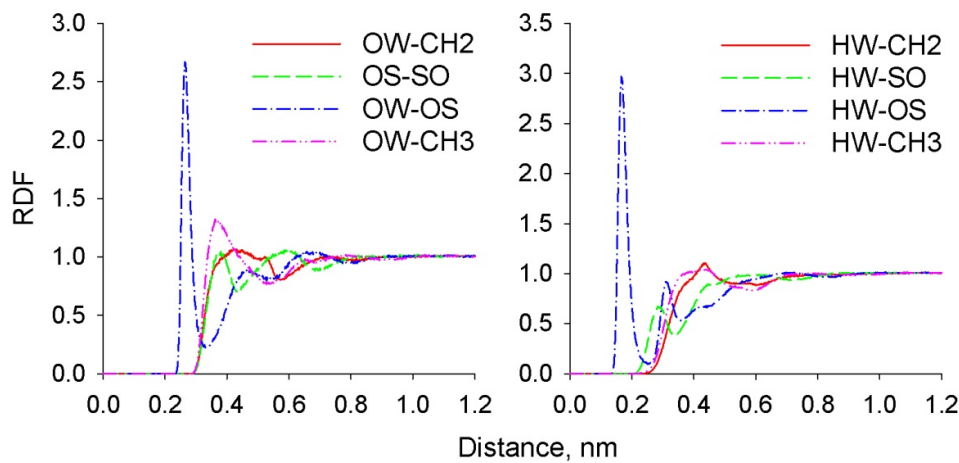

Figure 5. Radial distributions functions for selected interaction sites belonging to DESO and water in the 10 mol% DESO mixture with water at 300 K and 1 bar.

The water – sulfoxide hydrogen bond, identified in Figure 5, deserves a more detailed scrutiny. The position of the primary peak does not shift in response to the sulfoxide content (Figure 6). However, the height of the peak does change, from 3 points in 10% sulfoxide mixture to 7 points in 50% sulfoxide mixture. Interestingly, in EMSO the analogous hydrogen bonding peak is smaller (6.3 points).

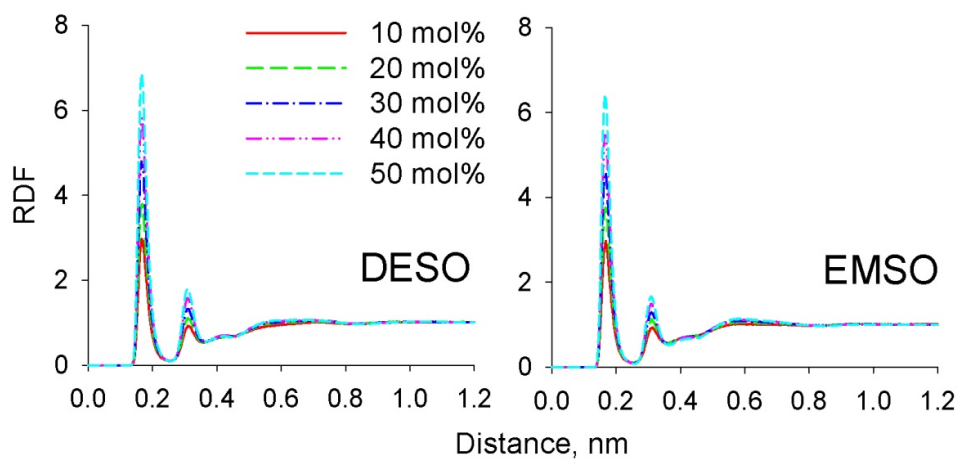

Figure 6. Radial distributions functions for the hydrogen bond centers (oxygen of the sulfoxide molecule and hydrogen of the water molecule) for all simulated mixtures. See the in-plot legend for line designation.

Running coordination numbers (Figure 7) are integrals of the radial distribution function computed from zero to the specified interatomic distance. Up to 0.3 nm, each DESO/EMSO molecule has only one neighbor water molecule, even in the water-rich mixtures. This result corresponds to the hydrogen bond, 0.19-0.21 nm, discussed above. The difference of mixture compositions is seen only beyond 0.3 nm, where the number of water molecules (the second hydration shell) is greater for lower sulfoxide and, thus, higher water concentrations.

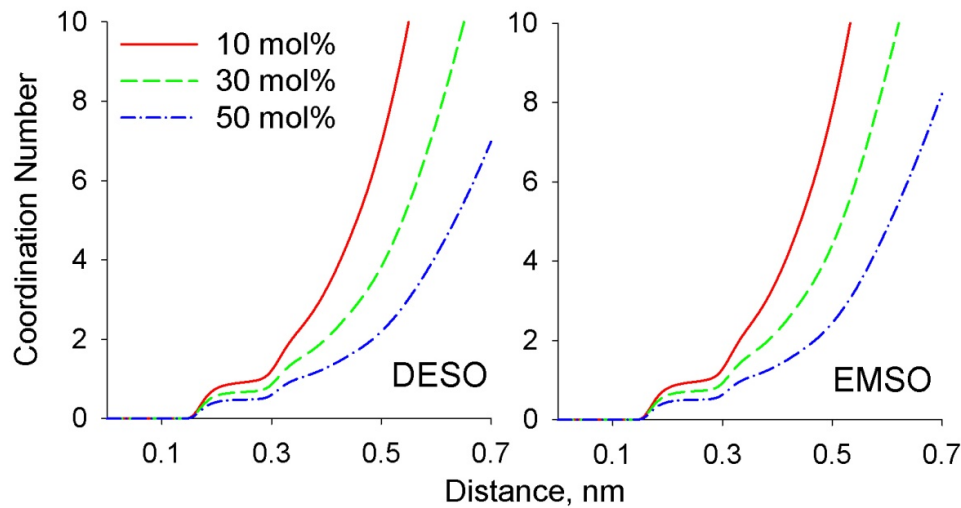

Figure 7. Running coordination numbers for water molecules around the DESO (EMSO) molecule, computed from the strongest pairwise interaction (Figure 6), in the 10, 20, and 30 mol% sulfoxide mixtures with water.

The oxygen-hydrogen non-covalent bond plays the most essential role in the negative deviation from ideality in the water-sulfoxide mixtures, whereas oxygen-hydrogen bonds in water maintain a unique network that positions normal boiling point of water so high. Figure 8 shows a number of hydrogen bonds in each 1 nm$^3$ of the mixtures. Whilst pure water system contains almost 60 hydrogen bonds, this number exponentially decays as the sulfoxide content increases. Decrease of the water molecules available in the vicinity of the sulfoxide molecule limits its possibility to get properly hydrated and ruin sulfoxide-sulfoxide associated structure patterns. Furthermore, decrease of the number of water molecules, per se, heavily decreases density of hydrogen bonds in the mixtures. The difference between the DESO- and EMSO-containing mixtures is not pronounced, indicating that side chains insignificantly contributes to miscibility.

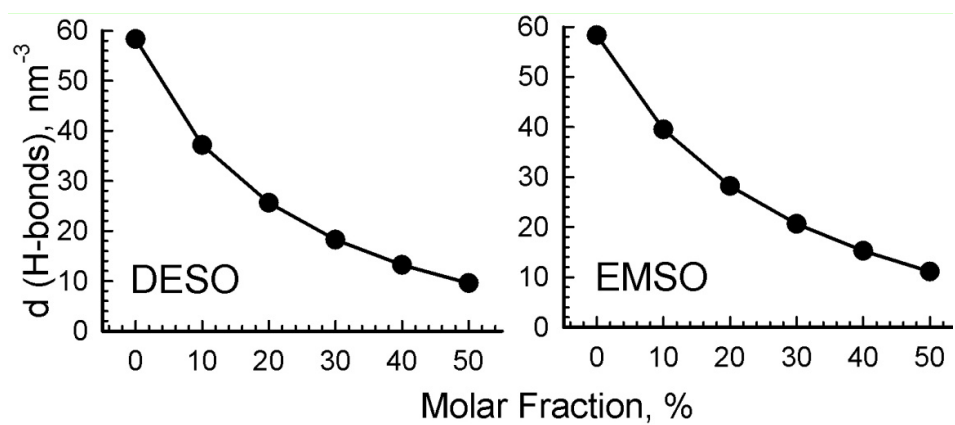

Figure 8. Normalized numbers of all hydrogen bonds in the DESO (right) and EMSO (left) mixtures with water as a function of the DESO (EMSO) molar fraction at 300 K and 1 bar. The size of symbols is comparable to the error bars computed.

Figure 9 provides only the numbers of sulfoxide – water hydrogen bonds. Expectedly, the density of hydrogen bonds is slightly larger in the EMSO-water equimolar mixture, because the volume of each EMSO molecule is slightly smaller than that of DESO molecule. This observation is valid for all mixture compositions. The fluctuations of the discussed property over time are quite modest, indirectly indicating a stable condensed-phase structure. In the 10%-sulfoxide mixtures, the number of sulfoxide-water hydrogen bonds per 1 nm$^3$ amounts to four, whilst this number increases up to 5.4 (DESO) and 6.7 (EMSO) for 30%-sulfoxide mixtures. Those mixtures exhibit the highest density of hydrogen bonds and likely the highest free energy gain due to mixing. Further increase of the sulfoxide molar content decreases hydrogen bonding. Analysis of the number of hydrogen bonds per unit volume vs. mixture composition is a valuable algorithm, because it allows to identify the most energetically (thermodynamically) favorable mixtures.

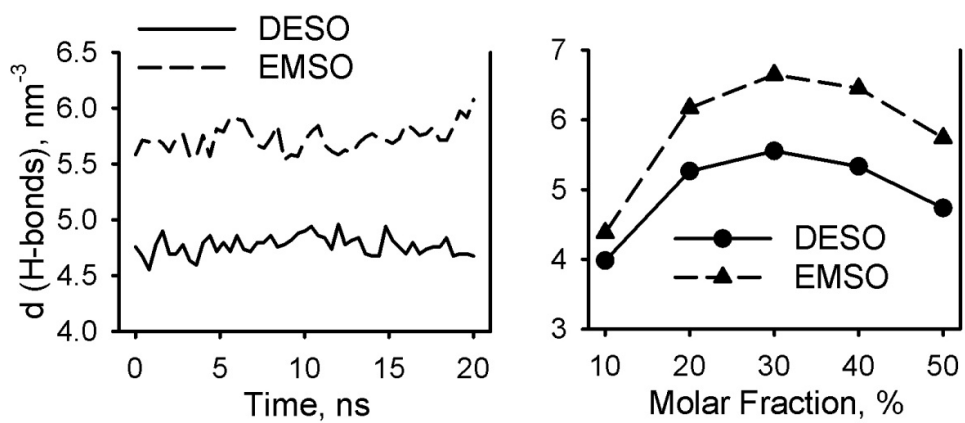

Figure 9. (Left) Evolution of the normalized numbers of the sulfoxide-water H-bonds in the equimolar mixtures of DESO (solid line) and EMSO (dashed line) with water. (Right) Average normalized numbers of the sulfoxide-water H-bonds vs. molar fraction of the sulfoxide in the mixture.

Sulfur-oxygen bonds can exhibit antiparallel alignment and, thus, give rise to sulfoxide clusters of different sizes (Figure 10). Clustering of sulfoxides is a reverse process to their mixing with water. The percentage of lone DESO molecules gradually decreases with an increase of the sulfoxide content in the mixture, from 80 to 70% of monomers. A large number of lone sulfoxide molecules indicate their excellent hydration and dispersion throughout an aqueous environment. The probability of finding a sulfoxide dimer is about 20%. The probabilities of a trimer and a tetramer do not exceed 5%. Qualitatively the same trends can be seen in the EMSO/water mixtures (Figure S1).

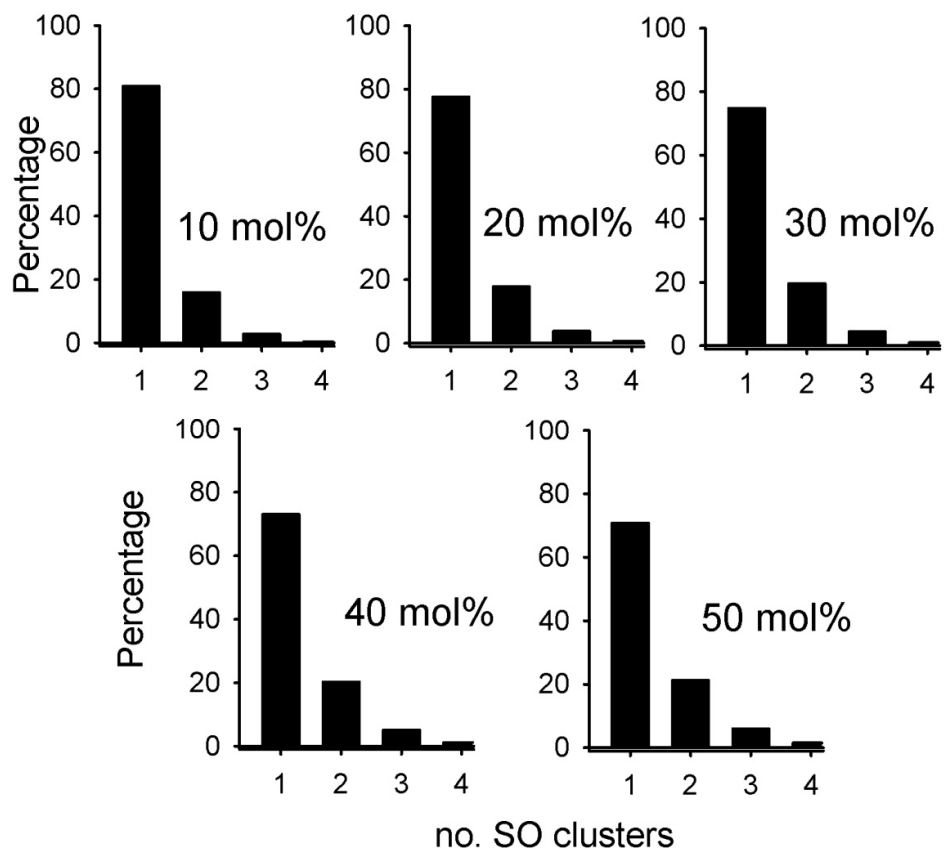

Figure 10. Probabilities of different cluster sizes of SO groups in the mixtures of DESO (10-50 mol%) with water. Clustering was performed using sulfur and oxygen atoms of DESO. The size of unity means a lone sulfoxide molecule. The distance threshold was taken from radial distribution functions.

Most sulfoxide molecules (up to 80%) exist as hydrated lone molecules (Figure 11). Upon increasing molar fraction of water, the percentage of lone sulfoxide molecules decreases slightly, whereas the percentages of dimers and trimers go upwards. The percentage of trimers is systematically larger than that of dimers, however, the difference between dimers and trimers is substantially smaller than the difference between dimers and monomers.

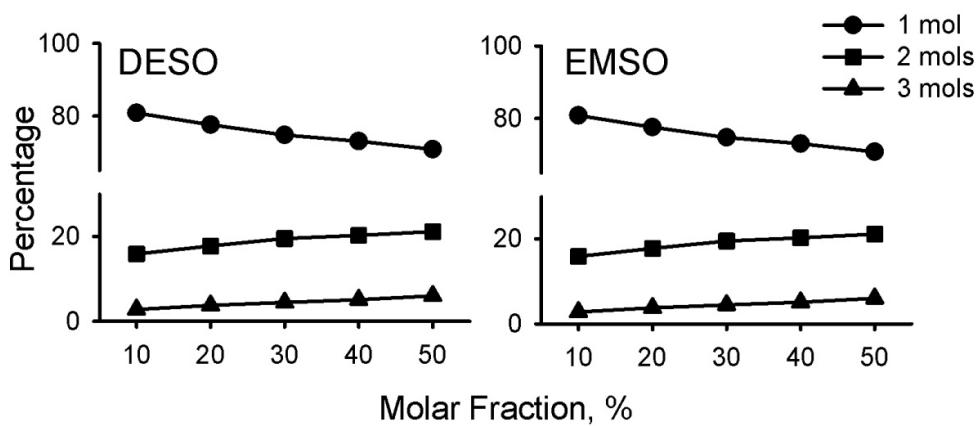

Figure 11. Probabilities of finding lone DESO (left) and lone EMSO (right) molecules and their respective small clusters (two and three tightly linked molecules) as a function of sulfoxide molar fraction. Clustering was performed using sulfur and oxygen atoms of DESO. The size of symbols is comparable to the error bars computed.

An average cluster size in the equimolecular DESO/water mixture is about 1.2 units (Figure 12). This result is in concordance with the above discussed distribution of sulfoxide molecules among clusters of different size. The fluctuations in time are insignificant, an amplitude being inferior to 0.1 units. In the meantime, the average maximum cluster size is about 4 DESO molecules. At certain points of time, a few DESO hexamers were detected as a statistically insignificant exception. A nearly harmonic fluctuation of cluster size distributions reveals that the system has attained proper equilibration.

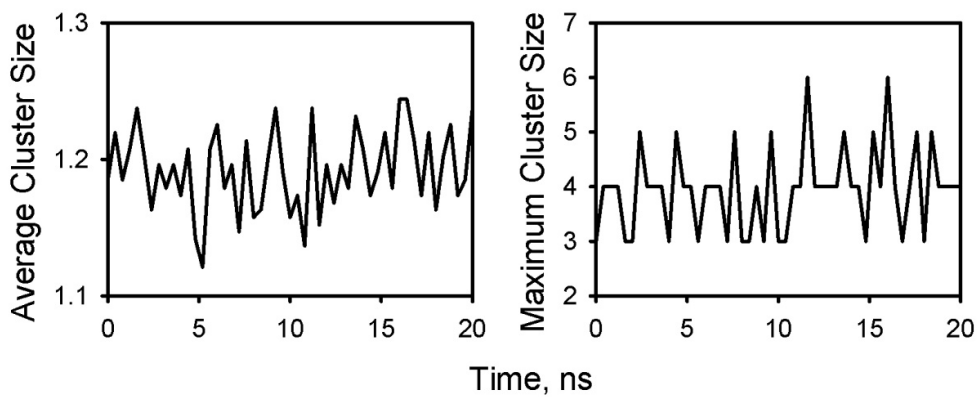

Figure 12. (Left) Average cluster size and (right) maximum cluster size in the DESO/water equimolar mixture at 300 K and 1 bar. Clustering was performed using sulfur and oxygen atoms of DESO.

Clustering can also be performed using all atoms belonging to the sulfoxide molecules. In this case, the cluster sizes are systematically larger. DESO forms large clusters, up to 50 molecules. Note, these so-called clusters are fairly unstable, being formed by thermal collisions in some time frames. The clusters of EMSO are smaller, because the volume of the molecule is somewhat smaller. Figure 13 additionally confirms that sulfoxide molecules are uniformly distributed throughout the simulation box (Figure 1). The cluster analyses reported herein show convincingly that self-association of EMSO and DESO is limited in their mixtures with water, being clearly inferior to the case of a more senior sulfoxide, DBSO.[19]

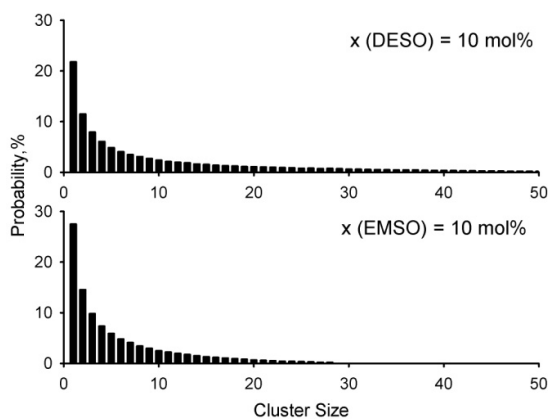

Figure 13. Cluster size distributions in the 10 mol% DESO (EMSO) mixtures. Clustering was performed using all atoms of DESO (EMSO).

The simulated self-diffusion constants were processed statistically using 10 independent pieces of trajectory for each mixture composition to obtain the averages and the error bars (Figures 14-15). The difference between results derived from independent trajectories is modest. Therefore, no sampling problems should be anticipated. Interestingly, water molecules slowdown DESO/EMSO molecules significantly, and vice versa, due to a strong electrostatic attraction, as revealed by radial distribution functions discussed above. EMSO is twice as mobile as DESO in pure liquid sulfoxides at 300 K, and the difference between them persists at all mixture compositions. It is rather noteworthy that a single methylene group alters self-diffusion so significantly in the liquid, whose structure is determined by the dipole-dipole interactions. Water molecules in the mixtures with DESO are slowdown somewhat more significantly, as compared to the EMSO-containing mixtures (Figure 15).

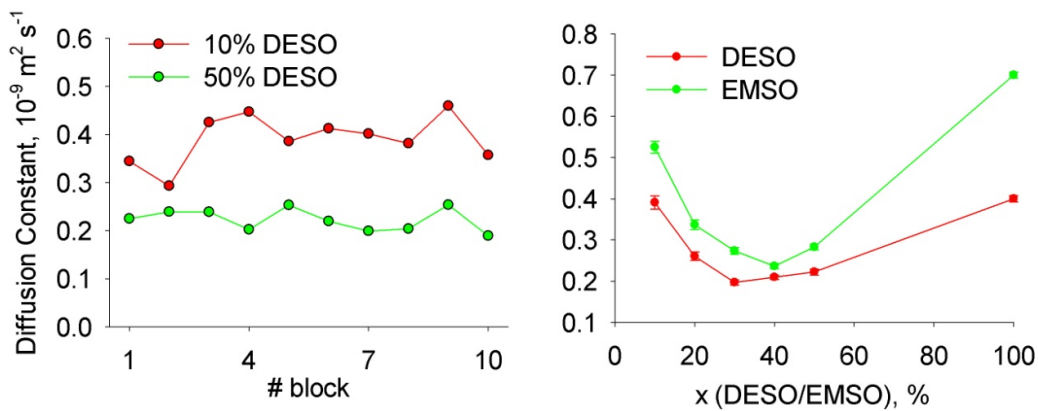

Figure 14. (Left) Block-averaging of self-diffusion coefficients using 10 independent parts of the computed phase trajectory. (Right) Self-diffusion coefficients of DESO and EMSO as a function of the sulfoxide molar fraction at 300 K and 1 bar. See legends for line designation.

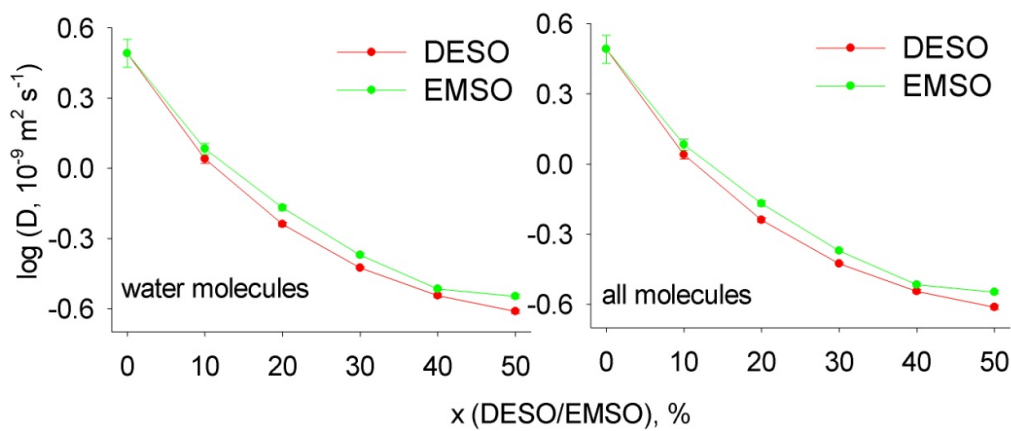

Figure 15. (Left) Self-diffusion coefficients of water molecules in their mixtures with DESO (EMSO). (Right) Average self-diffusion coefficients of the mixtures as a function of the sulfoxide molar fraction at 300 K and 1 bar.

Shear viscosity is generally in inverse proportion to self-diffusion. This is dictated by the Stokes-Einstein equation. Although this equation was first derived for spherical, one-atom molecules, it is often good for the molecules, whose shape differs for sphere quite significantly. In case of mixtures, an average diffusion coefficient in the system must be used. Indeed, since self-diffusion in the DESO/water mixtures is somewhat slower than that in the EMSO/water mixtures, the shear viscosities depict an inverse correlation (Figure 16). Importantly, the mixtures of both sulfoxides with water are more viscous than pure sulfoxides and by a few times more viscous than pure water at 300 K. The observed increased viscosity is in line with negative deviations from ideality in these mixtures and supports a conclusion that intermolecular interactions in the sulfoxide/water mixtures are stronger than in the pure components at the same physical conditions. The maximum shear viscosity in the EMSO/water mixtures is observed at 30 mol% of EMSO, whereas the maximum shear viscosity in the DESO/water mixtures is found at 40-50 mol% of DESO. The difference in particular values of shear viscosity is modest and almost fits within the error bars computed.

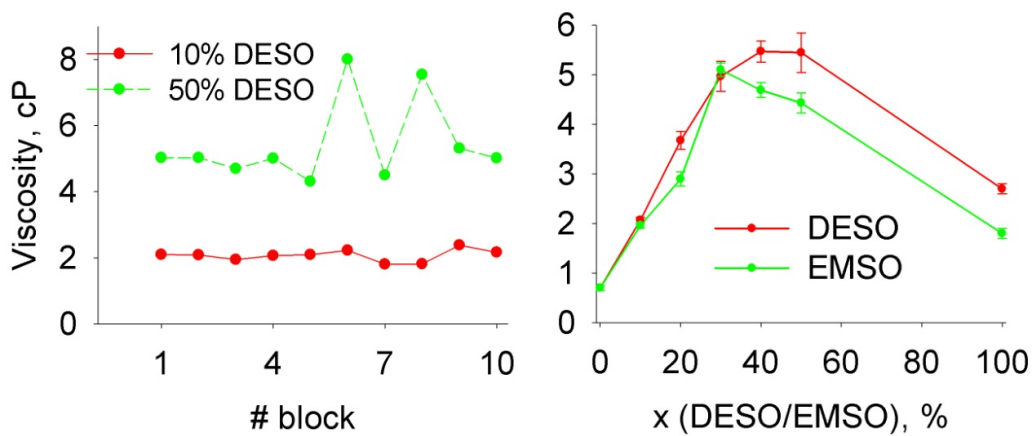

Figure 16. (Left) Block-averaging of shear viscosities of 10 and 50 mol% DESO-water mixtures. (Right) Shear viscosities of the DESO-water (red line) and EMSO-water (green line) mixtures as a function of their molar fractions. The shear viscosities were simulated at 300 K and 1 bar.

Gibbs free energies of solvation for EMSO/DESO in water and, respectively, water in EMSO/DESO were computed in order to estimate affinity of these liquids to one another and confirm the miscibility observed through the analysis of other properties. The calculation was performed by gradually decoupling (removing) of the solute molecule from its solvation environment, the algorithm, known as thermodynamic integration. An approximation of infinite dilution was used, i.e. a single solute molecule in a volume of a relatively large number of solvent molecules was simulated. It was found that Gibbs free energy of DESO solvation in water amounts to -40±1 kJ mol$^{-1}$, whilst solvation is insignificantly more favorable in case of EMSO, -41±1 kJ mol$^{-1}$. Expectedly, solvation of water in sulfoxides is somewhat less favorable, because sulfoxide do not possess so flexible network around the solute, as water molecules do. Compare, solvation of water is DESO amounts to -32±1 kJ mol$^{-1}$, whilst solvation of water in EMSO is -34±1 kJ mol$^{-1}$. EMSO possesses 3 hydrophobic groups, as compared to 4 groups in DESO, hence, miscibility of EMSO with water is more favorable. All reported solvation energies confirm outstanding mutual solvation of senior sulfoxides with water.

**Conclusions**

Classical molecular dynamics simulations of the DESO/water and EMSO/water liquid mixtures at 300 K and 1 bar using the recently developed force field were conducted. A variety of structure, thermodynamic, and dynamic physical chemical properties were computed and discussed over a wide composition range. Radial distribution function revealed a moderately strong hydrogen bonding between water and both sulfoxides, and confirmed antiparallel alignment of the S=O vectors. Cluster analysis provided a comprehensive description of how sulfoxide molecules are distributed throughout an aqueous phase and suggested existence of trimers and tetramers of the sulfoxides, whilst most their molecules are hydrated as monomers. Dynamic properties (self-diffusion and shear viscosity) are strongly influenced by one methylene group, that makes EMSO significantly more mobile that DESO and, furthermore, somewhat less viscous. Finally, Gibbs free energies of solvation/hydration are similar, with only insignificant advantage of EMSO over DESO. The results herein reported suggest that sulfoxide/water mixtures demonstrate interesting physical chemical properties and their superior performance as cryoprotectants and drug carriers must be expected and further validated.

**Supplementary Information**

Table S1 contains detailed distribution of cluster sizes in the EMSO/water mixtures.

**References**

(1) Bordat, P.; Sacristan, J.; Reith, D.; Girard, S.; Glattli, A.; Muller-Plathe, F., An improved dimethyl sulfoxide force field for molecular dynamics simulations. *Chem Phys Lett* **2003**, *374*, 201.
(2) Borin, I. A.; Skaf, M. S., Molecular association between water and dimethyl sulfoxide in solution: A molecular dynamics simulation study. *J Chem Phys* **1999**, *110*, 6412.
(3) Clark, T.; Murray, J. S.; Lane, P.; Politzer, P., Why are dimethyl sulfoxide and dimethyl sulfone such good solvents? *J Mol Model* **2008**, *14*, 689.
(4) Siddique, A. A.; Dixit, M. K.; Tembe, B. L., Solvation structure and dynamics of potassium chloride ion pair in dimethyl sulfoxide-water mixtures. *J Mol Liq* **2013**, *188*, 5.
(5) Patil, U. N.; Keshri, S.; Tembe, B. L., Solvation structure of sodium chloride (Na+-Cl-) ion pair in dimethyl sulfoxide-acetonitrile mixtures. *J Mol Liq* **2015**, *207*, 279.
(6) Qiao, X. S.; Zhao, T. X.; Guo, B.; Sha, F.; Zhang, F.; Xie, X. H.; Zhang, J. B.; Wei, X. H., Excess properties and spectral studies for binary system tri-ethylene glycol plus dimethyl sulfoxide. *J Mol Liq* **2015**, *212*, 187.
(7) Zhang, X. M.; Liu, H. P.; Liu, Y. X.; Jian, C. G.; Wang, W., Experimental isobaric vapor-liquid equilibrium for the binary and ternary systems with methanol, methyl acetate and dimethyl sulfoxide at 101.3 kPa. *Fluid Phase Equilibr* **2016**, *408*, 52.
(8) Zhang, N.; Li, W. Z.; Chen, C.; Zuo, J. G., Molecular dynamics simulation of aggregation in dimethyl sulfoxide-water binary mixture. *Comput Theor Chem* **2013**, *1017*, 126.
(9) Zhao, T. X.; Zhang, J. B.; Guo, B.; Zhang, F.; Sha, F.; Xie, X. H.; Wei, X. H., Density, viscosity and spectroscopic studies of the binary system of ethylene glycol plus dimethyl sulfoxide at T = (298.15 to 323.15) K. *J Mol Liq* **2015**, *207*, 315.
(10) Isobe, H.; Tanaka, T.; Nakanishi, W.; Lemiègre, L.; Nakamura, E., Regioselective Oxygenative Tetraamination of [60]Fullerene. Fullerene-mediated Reduction of Molecular Oxygen by Amine via Ground State Single Electron Transfer in Dimethyl Sulfoxide. *The Journal of Organic Chemistry* **2005**, *70*, 4826.
(11) Koley, S.; Kaur, H.; Ghosh, S., Probe dependent anomalies in the solvation dynamics of coumarin dyes in dimethyl sulfoxide-glycerol binary solvent: confirming the local environments are different for coumarin dyes. *Phys Chem Chem Phys* **2014**, *16*, 22352.

(12) Zhao, T. X.; Sha, F.; Xiao, J. B.; Xu, Q.; Xie, X. H.; Zhang, J. B.; Wei, X. H., Absorption, desorption and spectroscopic investigation of sulfur dioxide in the binary system ethylene glycol plus dimethyl sulfoxide. *Fluid Phase Equilibr* **2015**, *405*, 7.

(13) Markarian, S. A.; Gabrielyan, L. S., Dielectric relaxation study of ascorbic acid solutions in pure dimethylsulfoxide (or diethylsulfoxide) and in dimethylsulfoxide (or diethylsulfoxide)/water mixtures. *J Mol Liq* **2011**, *164*, 207.

(14) Markarian, S. A.; Gabrielyan, L. S.; Grigoryan, K. R., FT IR ATR study of molecular interactions in the urea/dimethyl sulfoxide and urea/diethyl sulfoxide binary systems. *J Solution Chem* **2004**, *33*, 1005.

(15) Slavic, M.; Djordjevic, A.; Radojicic, R.; Milovanovic, S.; Orescanin-Dusic, Z.; Rakocevic, Z.; Spasic, M. B.; Blagojevic, D., Fullerenol C-60(OH)(24) nanoparticles decrease relaxing effects of dimethyl sulfoxide on rat uterus spontaneous contraction. *J. Nanopart. Res.* **2013**, *15*.

(16) Gabrielyan, L.; Markarian, S.; Lunkenheimer, P.; Loidl, A., Low temperature dielectric relaxation study of aqueous solutions of diethylsulfoxide. *Eur Phys J Plus* **2014**, *129*.

(17) Markarian, S. A.; Gabrielian, L. S.; Bonora, S.; Fagnano, C., Vibrational spectra of diethylsulfoxide. *Spectrochim Acta A* **2003**, *59*, 575.

(18) Markarian, S. A.; Gabrielyan, L. S., Dielectric relaxation study of diethylsulfoxide/water mixtures. *Phys Chem Liq* **2009**, *47*, 311.

(19) Lo Celso, F.; Aoun, B.; Triolo, A.; Russina, O., Liquid structure of dibutyl sulfoxide. *Phys Chem Chem Phys* **2016**, *18*, 15980.

(20) Markarian, S. A.; Zatikyan, A. L.; Bonora, S.; Fagnano, C., Raman and FT IR ATR study of diethylsulfoxide/water mixtures. *J. Mol. Struct.* **2003**, *655*, 285.

(21) Markarian, S. A.; Zatikyan, A. L.; Grigoryan, V. V.; Grigoryan, G. S., Vapor pressures of pure diethyl sulfoxide from (298.15 to 318.15) K and vapor-liquid equilibria of binary mixtures of diethyl sulfoxide with water. *J Chem Eng Data* **2005**, *50*, 23.

(22) Markarian, S. A.; Bonora, S.; Bagramyan, K. A.; Arakelyan, V. B., Glass-forming property of the system diethyl sulphoxide/water and its cryoprotective action on Escherichia coli survival. *Cryobiology* **2004**, *49*, 1.

(23) Chaban, V. V., Vapor-liquid interface properties of diethyl sulfoxide-water and ethyl methyl sulfoxide-water mixtures: Molecular dynamics simulations and quantum-chemical calculations. *Fluid Phase Equilibr* **2016**, *427*, 180.

(24) Gabrielyan, L. S., FTIR and Ab Initio Studies of Diisopropylsulfoxide and its Solutions. *J. Solution Chem.* **2017**, *46*, 759.

(25) Bussi, G.; Donadio, D.; Parrinello, M., Canonical sampling through velocity rescaling. *J Chem Phys* **2007**, *126*, 014101.

(26) Parrinello, M.; Rahman, A., Polymorphic transitions in single crystals: A new molecular dynamics method. *J. Appl. Phys.* **1981**, *52*, 7182.

(27) Chaban, V. V., Force field development and simulations of senior dialkyl sulfoxides. *Phys Chem Chem Phys* **2016**, *18*, 10507.

(28) Chiu, S.-W.; Pandit, S. A.; Scott, H. L.; Jakobsson, E., An Improved United Atom Force Field for Simulation of Mixed Lipid Bilayers. *The Journal of Physical Chemistry B* **2009**, *113*, 2748.

(29) Darden, T.; York, D.; Pedersen, L., Particle mesh Ewald: An N·log(N) method for Ewald sums in large systems. *J. Chem. Phys.* **1993**, *98*, 10089.

(30) Berendsen, H. J. C.; van der Spoel, D.; van Drunen, R., GROMACS - A message-passing parallel molecular-dynamics implementation. *Computer Physics Communication* **1995**, *91*, 43.

(31) Van der Spoel, D.; Lindahl, E.; Hess, B.; Groenhof, G.; Mark, A. E.; Berendsen, H. J. C., GROMACS: Fast, flexible, and free. *J. Comput. Chem.* **2005**, *26*, 1701.

(32) Hess, B.; Kutzner, C.; van der Spoel, D.; Lindahl, E., GROMACS 4: Algorithms for Highly Efficient, Load-Balanced, and Scalable Molecular Simulation. *J Chem Theory Comput* **2008**, *4*, 435.

(33) Humphrey, W.; Dalke, A.; Schulten, K., VMD: Visual molecular dynamics. *J. Mol. Graphics* **1996**, *14*, 33.

(34) Martinez, L.; Andrade, R.; Birgin, E. G.; Martinez, J. M., PACKMOL: a package for building initial configurations for molecular dynamics simulations. *J. Comput. Chem.* **2009**, *30*, 2157.